\documentclass{ws-ijmpa}

\begin{document}

\markboth{Zimdahl, Velten \& Hip\'{o}lito-Ricaldi}
{Viscous Dark Fluid Universe}

%
\catchline{}{}{}{}{}
%

\title{VISCOUS DARK FLUID UNIVERSE: A UNIFIED MODEL OF THE DARK SECTOR?}

\author{WINFRIED ZIMDAHL}

\address{Universidade Federal do Esp\'{\i}rito Santo,
Departamento
de F\'{\i}sica,
Av. Fernando Ferrari, 514, Campus de Goiabeiras, CEP 29075-910,
Vit\'oria, Esp\'{\i}rito Santo, Brazil,
winfried.zimdahl@pq.cnpq.br}

\author{HERMANO E. S. VELTEN}

\address{Universidade Federal do Esp\'{\i}rito Santo,
Departamento
de F\'{\i}sica,
Av. Fernando Ferrari, 514, Campus de Goiabeiras, CEP 29075-910,
Vit\'oria, Esp\'{\i}rito Santo, Brazil, \\
velten@cce.ufes.br}

\author{WILIAM S. HIP\'{O}LITO-RICALDI}

\address{Universidade Federal do Esp\'{\i}rito Santo, Departamento de Ci\^encias Matem\'aticas e Naturais, CEUNES,
Rodovia BR 101 Norte, km. 60, CEP 29932-540,
S\~ao Mateus, Esp\'{\i}rito Santo, Brazil,
hipolito@ceunes.ufes.br}

\maketitle

\begin{history}
\received{Day Month Year}
\revised{Day Month Year}
\end{history}

\begin{abstract}
The Universe is modeled as consisting of pressureless baryonic matter
and a bulk viscous fluid which is supposed to represent a unified description of the dark sector.
In the homogeneous and isotropic background the \textit{total} energy density of this mixture behaves as a generalized Chaplygin gas. The perturbations of this energy density are intrinsically nonadiabatic and source relative entropy perturbations. The resulting baryonic matter power spectrum is shown to be compatible with the 2dFGRS and SDSS (DR7) data. A joint statistical analysis, using also Hubble-function and supernovae Ia data, shows that, different from other studies, there exists a maximum in the probability distribution for a negative present value of the deceleration parameter. Moreover, the unified model presented here favors a matter content that is of the order of the baryonic matter abundance suggested by big-bang nucleosynthesis. A problem of simple bulk viscous models, however, is the behavior of the gravitational potential and the reproduction of the CMB power spectrum.

\keywords{Structure formation; dark energy; dark matter.}
\end{abstract}

\ccode{PACS numbers: 98.80.-k, 95.35.+d, 95.36.+x, 98.65.Dx}

\section{Introduction}	

According to the prevailing interpretation, our Universe is dynamically dominated by a cosmological constant $\Lambda$ (or a dynamical equivalent, called dark energy (DE)) which contributes more than 70\% to the total cosmic energy budget. More than  20\% are contributed by cold dark matter (CDM) and only about 5\% are in the form of conventional, baryonic matter. Because of the cosmological constant problem in its different facets, including the coincidence problem, a great deal of work was devoted to alternative approaches in which a similar dynamics as that of the $\Lambda$CDM model is reproduced
with a time varying cosmological term, i.e., the cosmological constant is replaced by a dynamical quantity.
Both dark matter (DM) and DE manifest themselves so far only through their gravitational interaction. This provides a motivation for approaches in which DM and DE appear as different manifestation of one single
dark-sector component. The Chaplygin-gas model\cite{pasquier} and its
generalizations\cite{bertolami} realize this idea. Unified models of the dark sector of this type are attractive since one and the same component behaves as pressureless matter at high redshifts and as a cosmological constant in the long time limit. While the homogeneous and isotropic background dynamics for the (generalized) Chaplygin gas (GCG) is well compatible with the data, the study of the perturbation dynamics resulted in problems which apparently ruled out all Chaplygin-gas type models except those that are observationally almost indistinguishable from the $\Lambda$CDM model.\cite{Sandvik} To circumvent this problem, nonadiabatic perturbations were postulated and designed in a way to make the effective sound speed vanish.
But this amounts to an ad hoc procedure which leaves open the physical origin of nonadiabatic perturbations.
There exists, however, a different type of unified models of the dark sector, namely viscous cosmological models, which are intrinsically nonadiabatic.\cite{VDF}
In the homogeneous and isotropic background a one-component viscous fluid shares the same dynamics as a GCG.\cite{Szydlowski,BVM}  Now, what is observed in the redshift surveys is not the spectrum of the dark-matter
distribution but the baryonic matter spectrum. Including a baryon component into the perturbation dynamics for a universe with a Chaplygin-gas dark sector, there appears the new problem that the unified Chaplygin-gas scenario itself is disfavored by the data. It is only if the unified scenario with a fixed pressureless (supposedly) baryonic matter fraction of about $0.043$ (according to the results from WMAP and primordial nucleosynthesis) is \textit{imposed} on the dynamics, that consistency with the data is obtained.\cite{chaprel} If the pressureless matter fraction is left free, its best-fit value  is much larger than the baryonic fraction. In fact, it becomes even close to unity, leaving only a small percentage for the Chaplygin gas, thus invalidating the entire scenario. In other words, a Chaplygin-gas-based unified model of the dark sector is difficult to reconcile with observations. One may ask now, whether the status of unified models
can again be remedied by replacing the Chaplygin gas by a viscous fluid. It is this question that
we have investigated in Refs.~\refcite{VDF} and \refcite{BaVDF}, which are summarized in this contribution.
We studied the cosmological perturbation dynamics for a two-component model of baryons and
a viscous fluid, where the latter represents a one-component description of the dark sector.
We could show that such type of unified model is not only consistent for a fixed fraction of the baryons but also for the case that the matter fraction is left free.  Our analysis demonstrates that the statistically preferred value for the abundance of pressureless matter is compatible with the mentioned baryon fraction $0.043$ that follows from the synthesis of light elements. In addition we discuss here problems of the viscous model to correctly reproduce the anisotropy spectrum of the cosmic microwave background\cite{barrow} and give a brief outlook on a possible solution in the context of causal transport theory.\cite{oliver}

\section{The Two-Component Model}
The cosmic medium is assumed  to be describable  by an energy-momentum
tensor $T^{ik}$ which splits into a matter part $T^{ik}_{M}$ and viscous fluid part $T^{ik}_{V} $,
\begin{equation}
T^{ik} = \rho u^{i}u^{k} + p\left(g^{ik} +
u^{i}u^{k}\right), \qquad T^{ik} = T^{ik}_{M} + T^{ik}_{V}
 , \label{Tik}
\end{equation}
with
\begin{equation}
T^{ik}_{M} = \rho_{M} u^{i}_{M}u^{k}_{M} + p_{M}\left(g^{ik} +
u^{i}_{M}u^{k}_{M}\right),\quad T^{ik}_{V} = \rho_{V}
u^{i}_{V}u^{k}_{V} + p_{V}\left(g^{ik} + u^{i}_{V}u^{k}_{V}\right)
,\label{Tmv}
\end{equation}
where the subscript ``M" stands for matter and the subscript ``V" stands for viscous.
The total cosmic fluid is characterized by a four velocity $u^{m}$ while $u^{i}_{M}$ represents
the four velocity of the matter part and $u^{i}_{V}$ represents the four velocity of the viscous fluid.
Energy-momentum conservation is supposed to hold separately for
each of the components,
\begin{equation}
T^{ik}_{M\,;i} = T^{ik}_{V\,;i} = 0 \quad \Rightarrow\quad
T^{ik}_{\ ;i} = 0\ .\label{T;}
\end{equation}
Up to first order in the perturbations we have $\rho=\rho_M+\rho_V$ and $p=p_M+p_V$.
In general, the four velocities of the components are different. We
shall assume, however, that they coincide in the homogeneous and
isotropic zeroth order,
\begin{equation}
u^{i}_{M} = u^{i}_{V} = u^{i} \qquad \mathrm{(background )}  \ .
\label{}
\end{equation}
Difference will be important only at the perturbative level.
Let the matter be pressureless, i.e. $p_{M} = 0$ and the viscous fluid, according to Eckart's theory, be characterized by a bulk viscous pressure $p_{V} = p = - \zeta \Theta$,
where $\zeta = $ const and $\Theta = u^{i}_{;i}$ is the fluid
expansion. Under this condition the total pressure coincides with the pressure
of the viscous component. In terms of the present value $q_{0}$ of the deceleration parameter the Hubble rate  can be written as\cite{VDF}
\begin{equation}
\frac{H}{H_{0}} = \frac{1}{3}\,\left[1 - 2q_{0} + 2 \left(1 +
q_{0}\right)a^{-\frac{3}{2}}\right],
\label{r/r0q}
\end{equation}
which coincides with the Hubble rate of a specific ($\alpha = - 1/2$) GCG with the general equation of state
$p = -A/\rho^{\alpha}$.
Since $\rho_{M} =
\rho_{M0}a^{-3}$, we have
$\rho_{V} = \rho -
\rho_{M0}a^{-3}$.
It is the total energy density that behaves as a GCG, not the
component $V$. This type of unified model differs from unified models in which the total energy density
is the sum of a GCG and a baryon component. Only if the baryon component is ignored, both
descriptions coincide. Consequently, in the homogeneous and isotropic background,
a generalized Chaplygin gas with $\alpha = - 1/2$ can be seen as a
unified description of the cosmic medium, consisting of a separately conserved matter component and
a bulk viscous fluid with $\zeta = $ const, where the latter itself represents a unified model of the dark sector.

\section{Perturbation Dynamics}
\subsection{Total energy density perturbations}
The nonadiabaticity of the system as a whole
is characterized by
\begin{equation}
\frac{\hat{p}}{\rho + p} - \frac{\dot{p}}{\dot{\rho}}
\frac{\hat{\rho}}{\rho + p} = 3 H \frac{\dot{p}}{\dot{\rho}}
\left(\frac{\hat{\rho}}{\dot{\rho}} -
\frac{\hat{\Theta}}{\dot{\Theta}}\right),  \label{P-}
\end{equation}
where a caret denotes a perturbation quantity.
The expression (\ref{P-}) is governed by the dynamics of the total energy-density
perturbation $\hat{\rho}$ and by the perturbations $\hat{\Theta}$
of the expansion scalar, which is also a quantity that
characterizes the system as a whole. The behavior of these
quantities is described by the energy-momentum conservation for
the entire system and by the Raychaudhuri equation, respectively. Both of these
equations are coupled to each other.
The remarkable point is that these quantities and, consequently,
the total energy density perturbation, are independent of the
two-component structure of the medium. The reason is the direct relation $\hat{p} = -
\zeta\hat{\Theta}$
between the pressure perturbations and the
perturbations of the expansion scalar.
It is convenient to describe the perturbation dynamics in terms of
gauge invariant quantities which represent perturbations on
comoving hypersurfaces, indicated by a superscript $c$. These are defined as ($v$ is the velocity potential, defined by $\hat{u}_{\mu} = v_{,\mu}$)
\begin{equation}
\frac{\hat{\rho}^{c}}{\dot{\rho}} \equiv
\frac{\hat{\rho}}{\dot{\rho}} + v \ , \qquad
\delta \equiv \frac{\hat{\rho}^{c}}{\rho}
 \, .  \label{defc}
\end{equation}
For the fractional quantities we introduce the abbreviations
\begin{equation}
D^{c} \equiv \frac{\hat{\rho}^{c}}{\rho + p}\ ,\qquad P^{c} \equiv \frac{\hat{p}^{c}}{\rho + p}\ .\label{Dc}
\end{equation}
In our case we have
\begin{equation}
\frac{\hat{p}}{\dot{p}} = \frac{\hat{\Theta}}{\dot{\Theta}} \qquad
\Rightarrow\qquad \frac{\hat{p}^{c}}{\dot{p}} =
\frac{\hat{\Theta}^{c}}{\dot{\Theta}}
 .  \label{pcThetac}
\end{equation}
In terms of the comoving quantities the total energy and
momentum balances may be combined into (cf. Ref.~\refcite{VDF})
\begin{equation}
\dot{D}^{c} - 3H\,\frac{\dot{p}}{\dot{\rho}} \, D^{c}  +
\hat{\Theta}^c =0 \ . \label{dotD}
\end{equation}
The expansion scalar $\Theta $ is governed by the Raychaudhuri
equation, which, in linear order, can be
written in the form
\begin{equation}
\dot{\hat{\Theta}}^c + 2H\hat{\Theta}^c +
\frac{1}{a^{2}}\Delta P^{c} + \frac{3\gamma}{2}H^{2} \, D^{c}
= 0\ . \label{pertRay}
\end{equation}
It is through the
Raychaudhuri equation that the pressure gradient comes into play:
\begin{equation}
P^{c} = \frac{p}{\gamma \rho}\,\frac{\hat{\Theta}^{c}}{\Theta}\ ,
\quad \Rightarrow\quad
P^{c}  =
\frac{1}{2\gamma}\frac{p^{2}}{\rho^{2}} D^{c} - \frac{p}{3\gamma
\rho H}\dot{D}^{c}\ ,\label{PcDc}
\end{equation}
where $\gamma = 1 + \frac{p}{\rho}$.
The pressure perturbation consists of a term which is proportional
to the total energy-density perturbations $D^{c}$ (notice that the
factor in front of $D^{c}$ is positive), but additionally of a
term proportional to the time derivative $\dot{D}^{c}$ of $D^{c}$.
The relation between pressure perturbations $P^{c}$ and
energy perturbations $D^{c}$ is no longer simply algebraic, equivalent to a
(given) sound-speed parameter as a factor relating the two. The
relation between them becomes part of the dynamics. In a sense,
$P^{c}$ is no longer a ``local" function of $D^{c}$ but it
is a function of the derivative $\dot{D}^{c}$ as well\cite{essay}. This is equivalent to
$\hat{p} = \hat{p}(\hat{\rho}, \dot{\hat{\rho}})$. It is only for
the background pressure that the familiar dependence $p = p(\rho)$
is retained.
As already mentioned, the two-component structure of the medium is not relevant here.

Introducing now
\begin{equation}
\delta \equiv \gamma D^{c} =\frac{\hat{\rho}^{c}}{\rho} \,
\ ,  \label{deltac}
\end{equation}
and changing from the variable  $t$ to $a$,
Eqs.~(\ref{dotD}) and (\ref{pertRay}) may be combined to yield the second-order equation
\begin{equation}
\delta'' + f\left(a\right)\delta' + g\left(a\right) \,\delta = 0 \ ,\label{dddshort}
\end{equation}
where $\delta' \equiv \frac{d \delta}{d a}$ and the coefficients $f$ and $g$ are
\begin{equation}
f\left(a\right) = \frac{1}{a}\,\left[\frac{3}{2} - 6\frac{p}{\rho}  - \frac{1}{3}\frac{p}{\gamma\rho}\,\frac{k^{2}}{H^{2}
a^{2}}\right]  \label{f}
\end{equation}
and
\begin{equation}
g\left(a\right) = - \frac{1}{a^{2}}\,\left[\frac{3}{2}  + \frac{15}{2}
\frac{p}{\rho} - \frac{9}{2}\,\frac{p^{2}}{\rho^{2}}
- \frac{1}{\gamma}\frac{p^{2}}{\rho^{2}} \frac{k^{2}}{H^{2}
a^{2}}\right]\ , \label{g}
\end{equation}
respectively.
Equation (\ref{dddshort}) coincides with the corresponding equation for the one-component case in Ref.~\refcite{VDF}.

\begin{figure}  
\hspace{0cm}
\includegraphics[width=12cm, height=13cm]{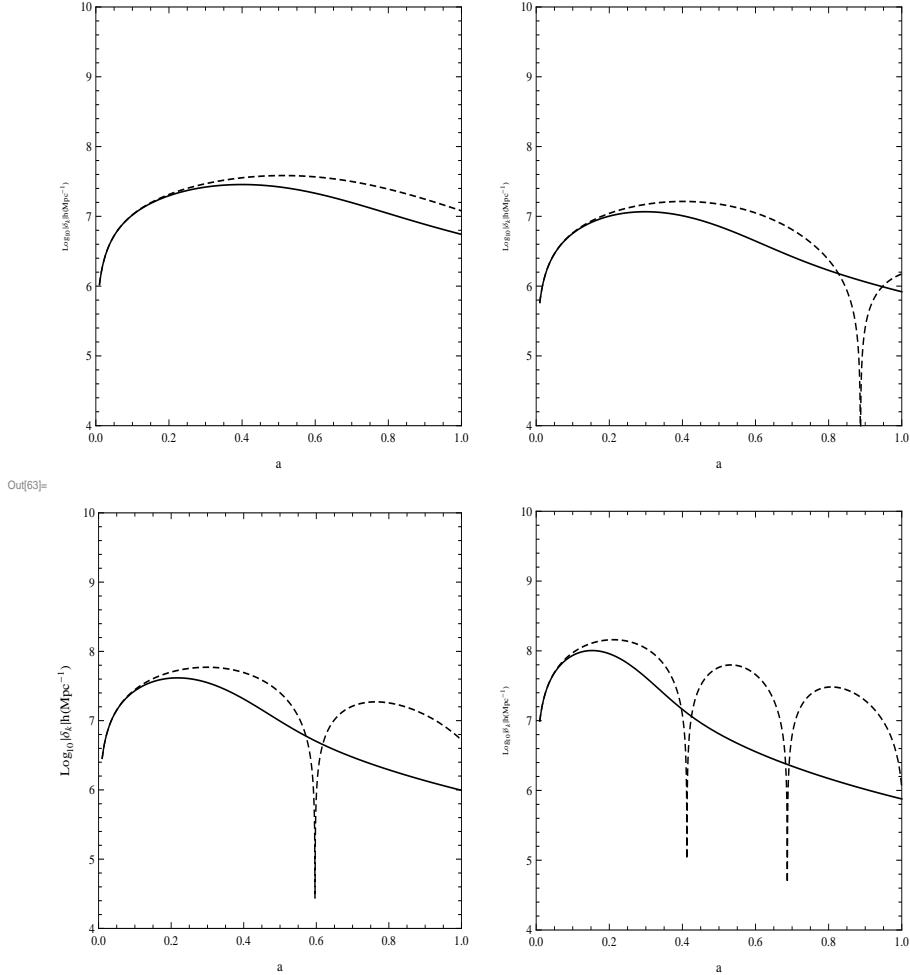}
\vspace*{8pt}
\caption{\label{fig2}
Absolute values (logarithmic scale) of density fluctuations as function of the scale factor $a$
for $\nu=-1$ ($\alpha=1/2$) and $q_0=-0.5$  for different
scales. The values of $k$ are $k=0.5$ (top left), $k= 0.7$ (top right), $k =1$ (bottom left)
and $k =1.5$ (bottom right), all in units of $h Mpc^{-1}$. Solid curves represent the bulk viscous
model, dashed curves the corresponding GCG model.
Notice that both models are always different, except
at very early times.}
\end{figure}

In Fig.~\ref{fig2} the density fluctuations for the viscous model are compared with those of the GCG model for different values of the relevant parameters. Although identical in the background, both models are qualitatively very different at the perturbative level.
The density perturbations in the bulk-viscous scenario are  well behaved
at all times. The GCG model predicts (unobserved) oscillations, as was also found in Ref.~\refcite{ioav}. The latter behavior was the main reason for discarding these models, except, possibly, for very small values of $\alpha$.
This unwanted property does not hold for our viscous model. This coincides with the results of Ref.~\refcite{BVM}.
Both models coincide for early times, confirming our previous analytical result, that non-adiabatic contributions are negligible in the past, but become relevant at a later period. The non-adiabatic contributions
are essential to avoid the mentioned unrealistic features of  GCG models.

\subsection{Relative entropy perturbations}
The relative entropy perturbations are defined by
\begin{equation}
S_{MV} \equiv \frac{\hat{\rho}_{M}}{\rho_{M}} -
\frac{\hat{\rho}_{V}}{\rho_{V} + p_{V}}  \ \label{SVM}
\end{equation}
and obey the equation
\begin{equation}
S_{VM}'' + r(a)S_{VM}' + s(a) S_{VM} = c(a) \delta' + d(a)\delta
 \
\label{ppS}
\end{equation}
with the coefficients
\begin{equation}
r(a) = \frac{1}{a}\left[\frac{3}{2} - \frac{3}{2}\frac{p}{\rho}  -
3\frac{p}{\rho}\frac{\rho_{M}}{\rho_{V} +
p}\right]
 \ ,
\label{r(a)}
\end{equation}
\begin{equation}
s(a) = - \frac{3}{a^{2}}
\frac{p}{\rho}\frac{\rho_{M}}{\rho_{V} + p}\left[1 +
\frac{3}{4}\frac{p}{\rho}\right]
 \ ,
\label{s(a)}
\end{equation}
\begin{equation}
c(a) = \frac{1}{a}\left[\frac{3}{\gamma}\frac{p}{\rho_{V} +
p}\,\left(1 +
\frac{p}{2 \rho} + \left(1 + \frac{p}{\gamma
\rho}\right)\frac{k^{2}}{9 H^{2}a^{2}}\right)\right]
 \
\label{c(a)}
\end{equation}
and
\begin{equation}
d(a) = \frac{9}{2\gamma a^{2}}\frac{p}{\rho_{V} +
p}\,\left[\left(1
-\frac{p}{\rho}\right)\left(1 + \frac{p}{2 \rho}\right) -
2\frac{p}{\rho}\left(1 + \frac{p}{\gamma
\rho}\right)\frac{k^{2}}{9
H^{2}a^{2}}\right]
 \ .
\label{d(a)}
\end{equation}
\ \\
The set of equations (\ref{dddshort}) and (\ref{ppS}) contains the
entire perturbation dynamics of the system. At first, the homogeneous Eq.~(\ref{dddshort}) for $\delta$ has to be solved. Subsequently, once $\delta$ is known, Eq.~(\ref{ppS}) determines the relative entropy perturbations.

\subsection{Baryonic perturbations}

The quantity relevant for the observations is the
fractional perturbation $\delta_{M} \equiv \frac{\hat{\rho}_{M}^{c}}{\rho_{M}}$ of the energy density of the baryons, given by
\begin{equation}
\delta_{M} = \frac{1}{\gamma}\left[\delta - \frac{\rho_{V} +
p}{\rho}S_{VM}\right] \ .
 \label{del1}
\end{equation}
At early times, i.e. for small scale factors $a \ll 1$, the equation (\ref{dddshort}) has the asymptotic form
\begin{eqnarray}\label{EarlyViscous}
\qquad \qquad \delta'' + \frac{3}{2a} \,\delta' - \frac{3}{2a^2}\,\delta =0 \,,\qquad \qquad  (a \ll 1)
\end{eqnarray}
independent of $q_0$ and for all scales.
The solutions of (\ref{EarlyViscous}) are
\begin{eqnarray} \label{asympsol}
\delta(a\ll 1)= c_1 a + c_2 a^{-3/2} \,,
\end{eqnarray}
where $c_1$ and $c_2$ are integration constants.
The nonadiabatic contributions to the total density perturbations are negligible at high redshifts.\cite{VDF}
For $a \ll 1$ the coefficients $s(a)$, $c(a)$ and $d(a)$ in (\ref{ppS}) become negligible and
$r(a)\rightarrow \frac{3}{2}$. Eq.~(\ref{ppS}) then reduces to
\begin{equation}
S_{VM}'' + \frac{3}{2a}S_{VM}'  = 0\,,\qquad \qquad  (a \ll 1)
 \ .
\label{ppS1}
\end{equation}
It has the solution $S_{VM} = $ const $=0$. From the definition (\ref{SVM}) we find that at high redshifts
\begin{equation}
S_{MV} = \frac{\hat{\rho}_{M}}{\rho_{M}} -
\frac{\hat{\rho}_{V}}{\rho_{V}}\,,\qquad \qquad  (a \ll 1)  \ .\label{SVM1}
\end{equation}
Consequently, both the nonadiabatic contributions to the total energy-density fluctuations and the
relative entropy perturbations are negligible and we have almost purely adiabatic perturbations $\delta_{M} = \delta$ at $a \ll 1$.
This allows us to relate our model to the $\Lambda$CDM model at early times.
We shall use the fact that the matter power spectrum for the
$\Lambda$CDM model is well fitted by the BBKS transfer
function.\cite{bbks}
Integrating the $\Lambda$CDM model back from today to a distant past, say $z = 1.000$, we
obtain the shape of the transfer function at that moment.
The
spectrum determined in this way is then used as initial
condition for our viscous model. For more details see Ref.~\refcite{juliosola}.

\section{Statistical Analysis}
\label{Numerical analysis}

To estimate the free parameters of our model we perform a Bayesian analysis and construct the corresponding
probability distribution functions. At first we consider large-scale-structure data from the 2dFGRS\cite{cole} and
SDSS DR\cite{sdss} programs.
The matter power spectrum is defined by
\begin{equation}
P_k=\left|\delta_{M,k}\right|^{2}\ ,
\end{equation}
where $\delta_{M,k}$ is the Fourier component of the density contrast $\delta_{M}$.
Generally, for a set of free parameters $\left\{{\bf p}\right\}$, the agreement between the theoretical prediction and observations is assessed by minimizing the quantity
\begin{equation}
\chi^{2}\left({\bf p}\right)=\frac{1}{Nf}\sum_{i}\frac{\left[P^{th}_{i}({\bf p}) - P^{obs}_{i}({\bf p})\right]^{2}}{\sigma_{i}^{2}},
\end{equation}
where Nf means the number of degrees of freedom in the analysis. The quantities $P^{th}_{i}$ and $P^{obs}_{i}$ are the theoretical and the observed values, respectively, of the power spectrum and $\sigma_{i}$ denotes the error for the data point $i$. With the help of $\chi^2$ we then construct the probability density function (PDF)
\begin{equation}
P = \mathcal{B} \,e^{-\frac{\chi^{2}(\bf{p})}{2}}\ ,
\end{equation}
where $\mathcal{B}$ is a normalization constant.

\begin{center}
\begin{figure}[!h]
\hspace{0cm}
\begin{minipage}[t]{0.45\linewidth}
\includegraphics[width=\linewidth]{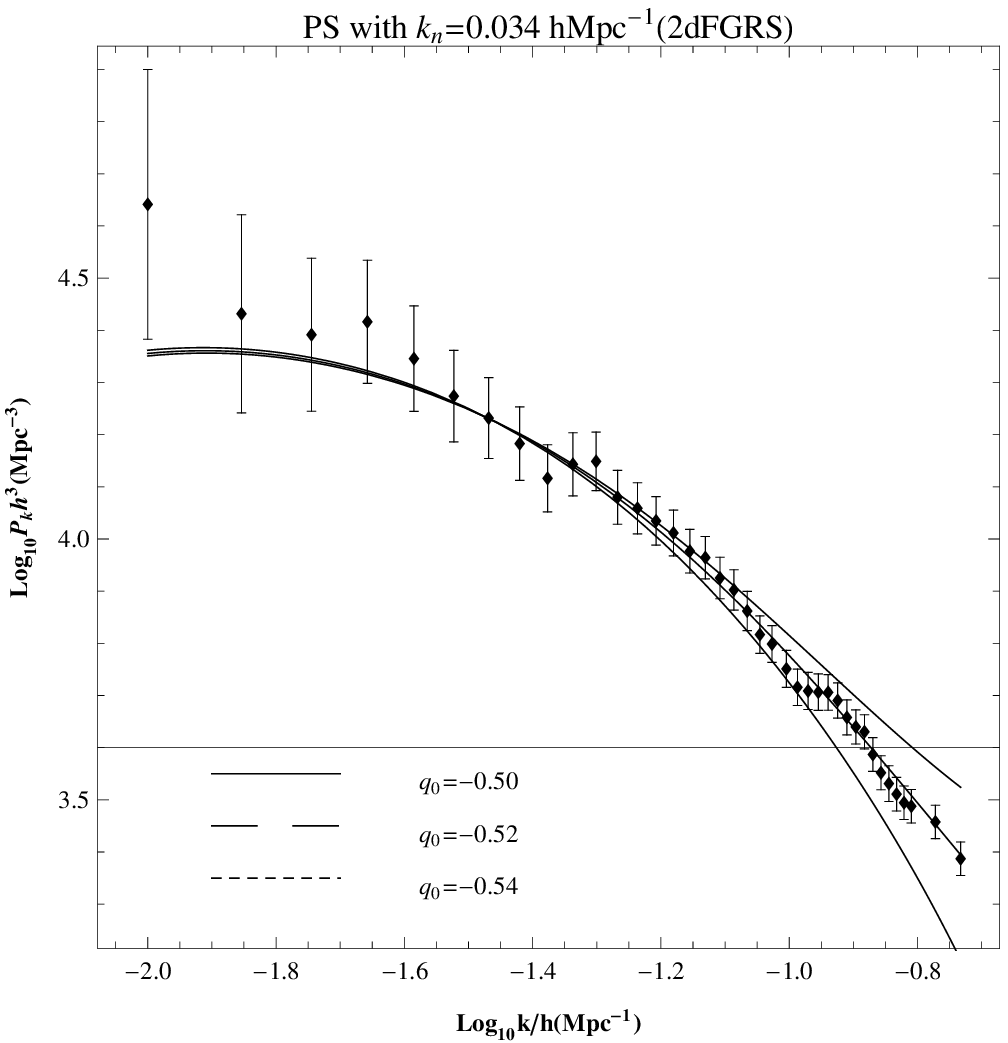}
\end{minipage} \hfill
\begin{minipage}[t]{0.45\linewidth}
\includegraphics[width=\linewidth]{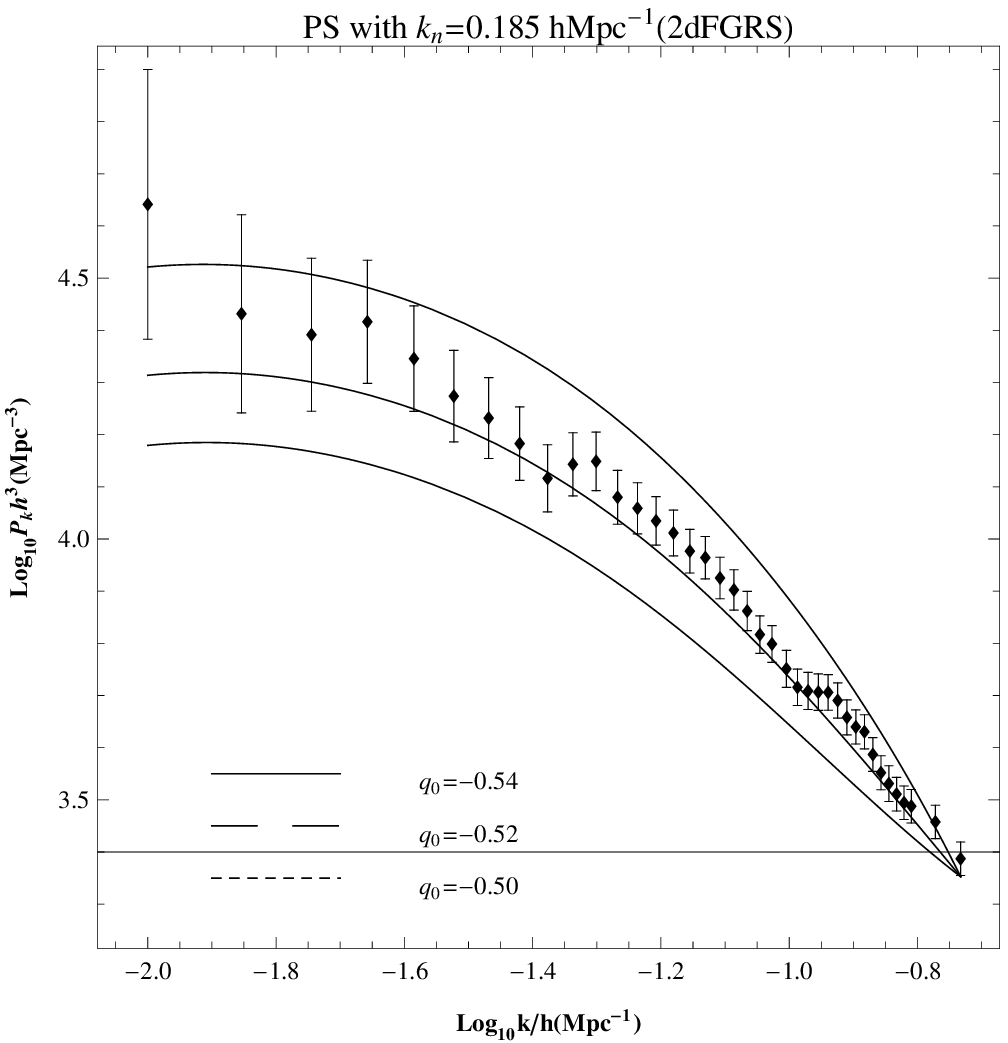}
\end{minipage} \hfill
\begin{minipage}[t]{0.45\linewidth}
\includegraphics[width=\linewidth]{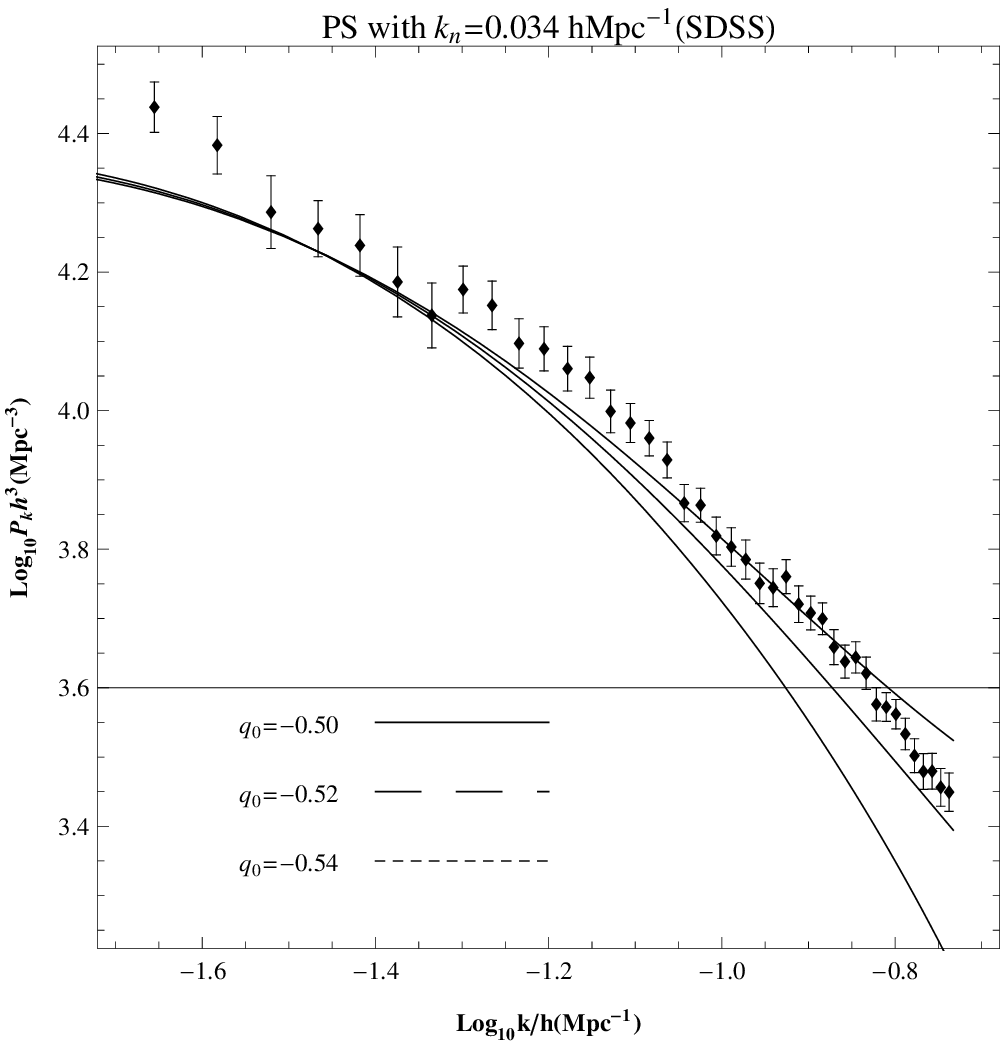}
\end{minipage} \hfill
\begin{minipage}[t]{0.45\linewidth}
\includegraphics[width=\linewidth]{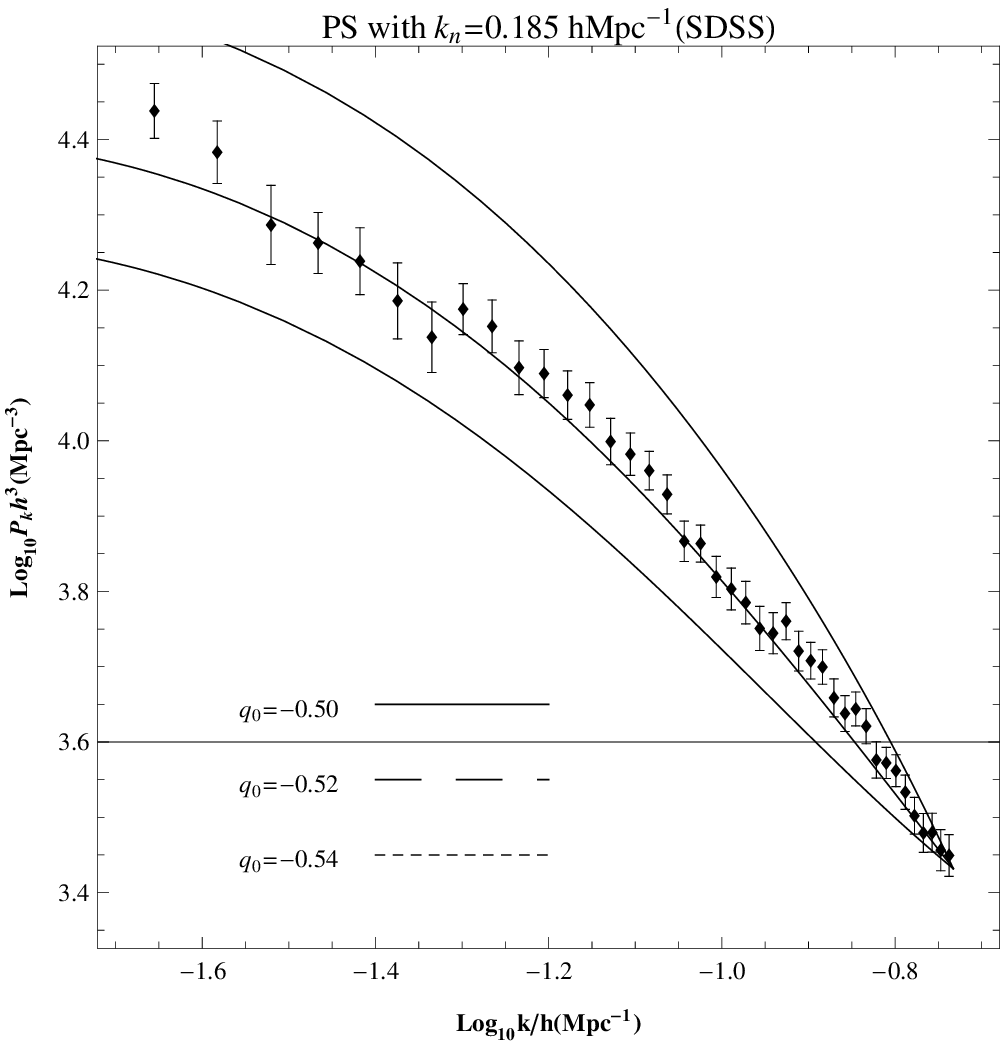}
\end{minipage} \hfill
\caption{{\protect\footnotesize Power spectra (PS) normalized at $k_{n}=0.034 hMpc^{-1}$ (left panels) and at $0.185 hMpc^{-1}$ (right panels) for different negative values of $q_{0}$. The top panels compare the PS with the 2dFGRS data, the bottom panels with the SDSS DR7 data.}}
\label{2}
\end{figure}
\end{center}

\begin{center}
\begin{figure}[!h]
\hspace{0cm}
\begin{minipage}[t]{0.4\linewidth}
\includegraphics[width=\linewidth]{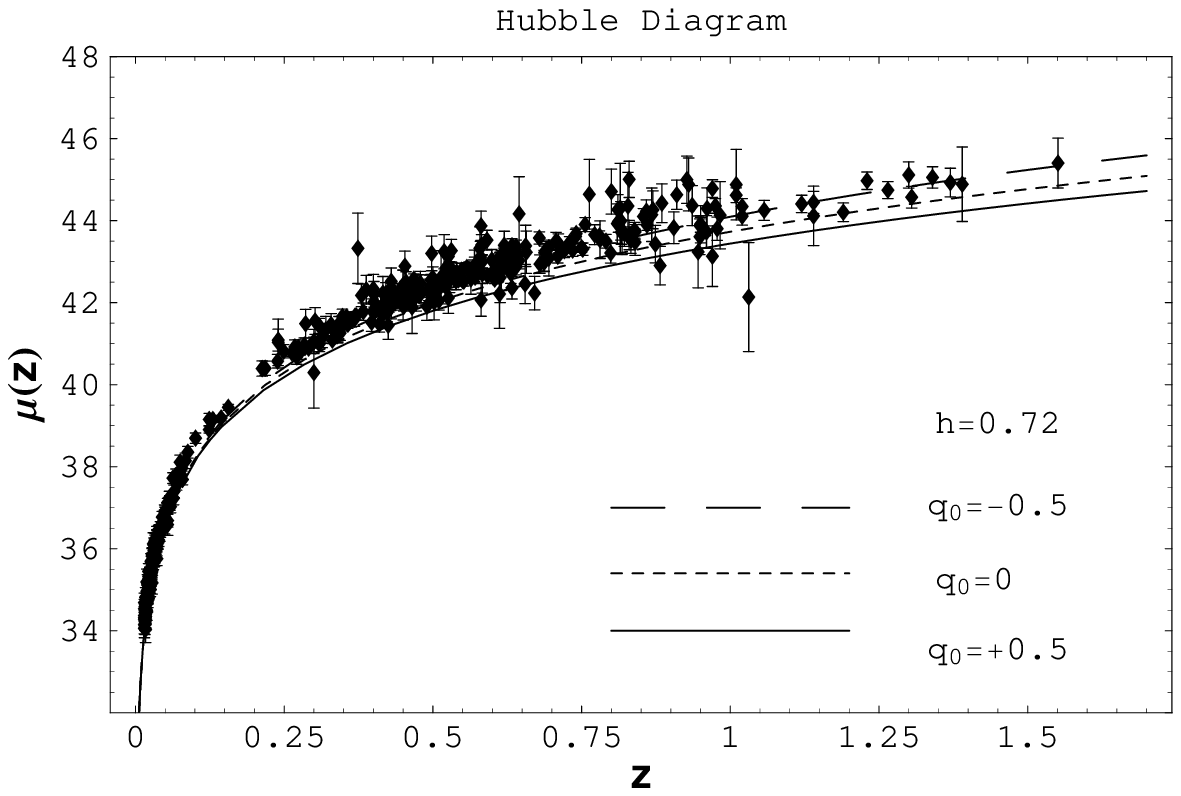}
\end{minipage} \hfill
\begin{minipage}[t]{0.25\linewidth}
\includegraphics[width=\linewidth]{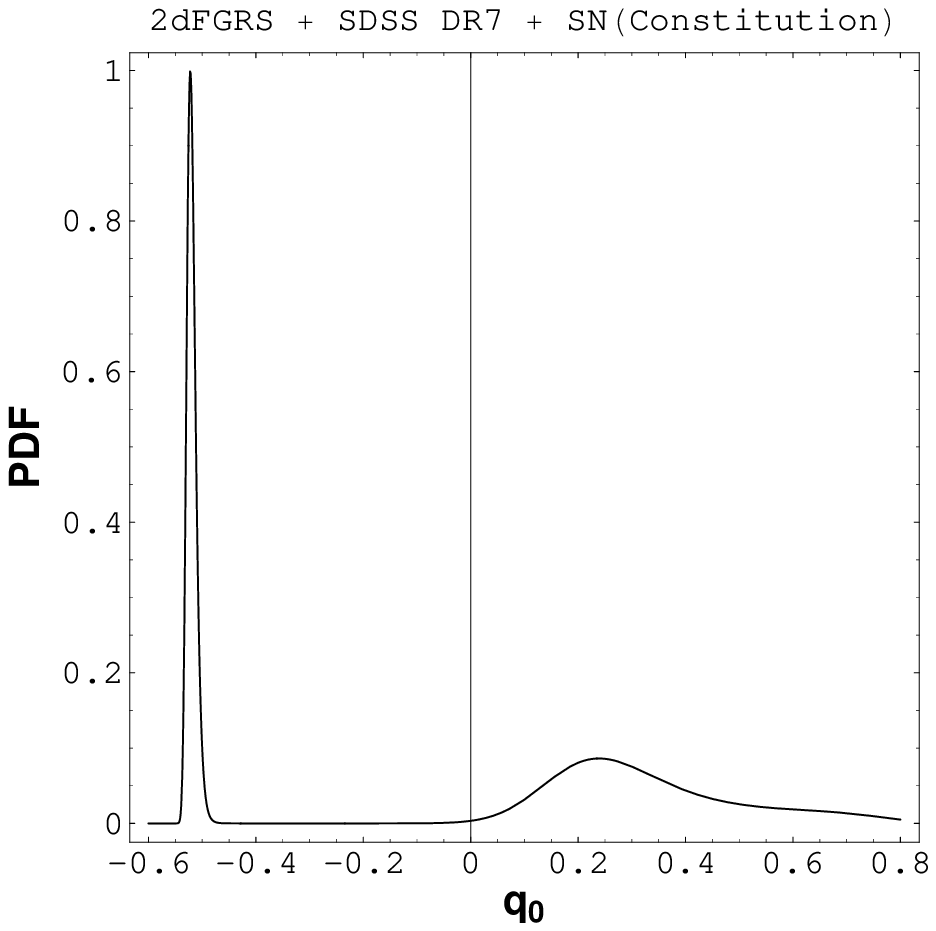}
\end{minipage} \hfill
\begin{minipage}[t]{0.26\linewidth}
\includegraphics[width=\linewidth]{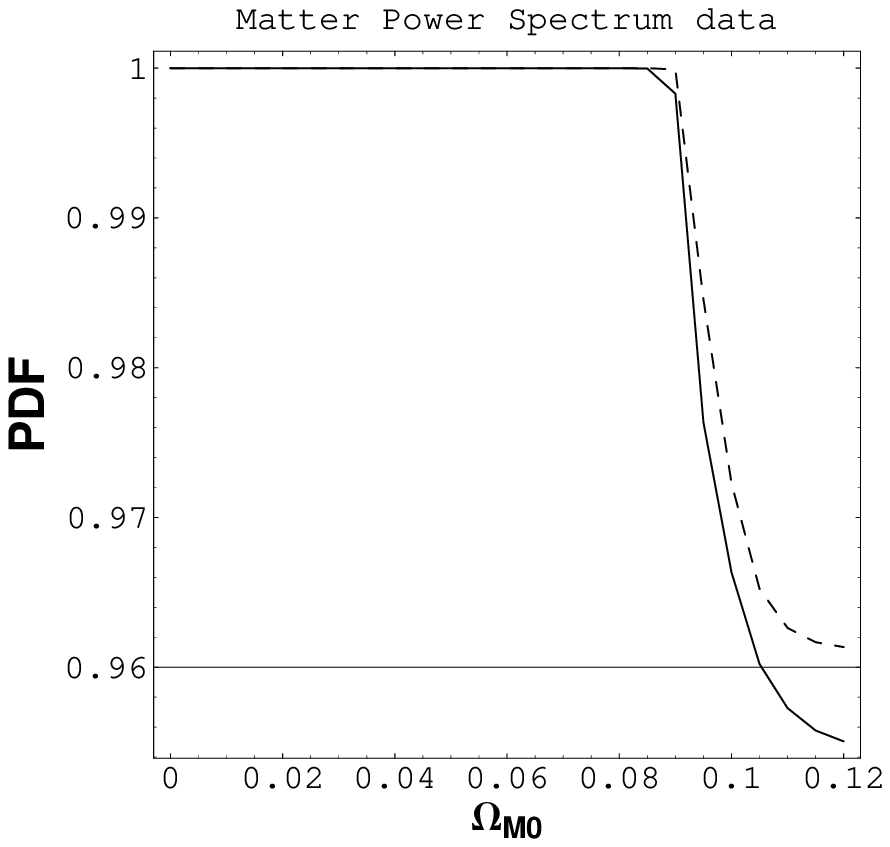}
\end{minipage} \hfill
\vspace*{8pt}
\caption{Left panel: Hubble diagram, center panel:
probability distribution function (PDF) for $q_{0}$, based on a joint
analysis of matter power spectrum and SN Ia data. The right panel shows
the PDF for the pressureless component.}
\label{SN}
\end{figure}
\end{center}

To test our model against the observed power-spectra data we consider the following two situations.
(i) We assume the matter component to be entirely baryonic with a fraction $\Omega_{M0} = 0.043$ as suggested by the WMAP data. Fixing also $H_{0} = 72$, a value favored by these data as well, the only remaining free parameter is $q_{0}$. This will provide us with information about the preferred value(s) of $q_{0}$ for the unified dark-sector model.
Fig.~\ref{2} shows the theoretically obtained spectrum for various values of $q_{0}$ together with the power-spectrum data points.
To better illustrate the relation between the predictions of the model and the observations, two different normalization wave numbers, $k_{n}=0.034hMpc^{-1}$ and $k_{n}=0.185hMpc^{-1}$, have been chosen, but our statistical results do not depend on a specific normalization.
(ii) We leave the matter fraction free, thus admitting that the matter component is not only made up by the baryons. This is equivalent to allow for a separate DM component in addition to the contribution effectively  accounted for by the viscous fluid.
This additional freedom is used to test our unified model of the dark sector itself. The unified model can be regarded as favored by the data if the PDF for the matter fraction is large around the value that characterizes the baryon fraction. If, on the other hand, the PDF is largest at a substantially higher value, the unified model has to be regarded as disfavored. The results of our statistical analysis are visualized in Fig.~\ref{SN}. The center panel shows a maximum of the PDF for $q_{0}$ at $q_{0}\approx -0.53$. According to the right panel the matter-fraction probability is highest for
$\Omega_{M0}\lesssim 0.08$ and sharply decays for $\Omega_{M0} > 0.08$.

\section{Comments}
\label{comments}
Before summarizing our results, two comments are in order here. The first one concerns a more adequate, ``causal"  description of the viscous fluid, the second one points out the possibility to mimic unified models of the dark sector in terms of kinematic back reactions in an averaged inhomogeneous cosmology.

\subsection{Viscous fluid and gravitational potential}

While the analysis of the matter power spectrum data favors the viscous model, the situation changes if the spectrum of the CMB is considered. A recent investigation showed, that, at least for the Eckart theory applied here, the time dependence of the gravitational potential differs dramatically from the corresponding dependence of the $\Lambda$CDM model.\cite{oliver} This confirms an earlier analysis in Ref.~\refcite{barrow} and severely weakens the status of a viscous dark-sector model. Now, it is well known, that Eckart's theory suffers from causality and stability problems. Therefore, one may hope that a more adequate description of the dark sector on the basis of causal thermodynamics can cure this shortcoming. The essential difference to Eckart's theory is, that the algebraic relation $\Pi = - \zeta \Theta$ which we used in our analysis, is replaced by a differential equation
\begin{equation}
\Pi+ \tau\dot{\Pi}=-\zeta \Theta \ ,
\end{equation}
where the relaxation time $\tau$ appears as a new parameter. This parameter is related to the propagation speed of viscous pulses. It could be demonstrated, that with a very small value of this dissipative sound speed of the order $\lesssim 10^{-8}$, this dissipative dark-sector model could indeed produce a gravitational potential similar to that of the $\Lambda$CDM model.\cite{oliver}

\subsection{Kinematic back reaction and unified models }
There exists a line of research which tries to explain the accelerated expansion of the Universe as the consequence of a back reaction of a suitably averaged inhomogeneous cosmology. This approach, which avoids the introduction of dark energy, relies on the fact that the averaged Einstein tensor on the left-hand side of the field equations is different from the Einstein tensor of an averaged metric. This difference results in kinematic back-reaction terms, which modify the standard background dynamics.\cite{buchert} Effective fluid models of this back reaction have been constructed which include the Chaplygin gas.\cite{buchertchap} This seems to provide additional motivation for the further investigation of unified dark sector models in a different context.

\section{Conclusions}
\label{conclusions}
We have modeled the cosmic substratum at the present time as a mixture of a viscous fluid and baryons.
The viscous fluid is assumed to provide a  unified description of the cosmological dark sector.
In the homogeneous and isotropic background the two-component system of a bulk viscous fluid and a separately conserved baryon component behaves as a generalized Chaplygin gas.  The total energy-density perturbations, however, are intrinsically nonadiabatic and coincide with those of a one-component viscous fluid.
The fluctuations of the baryon component are obtained from a combination of the total energy density perturbations and relative entropy perturbations in the two-component system where the former source the latter.
The observed matter-power spectrum  is well reproduced. There do not appear oscillations or instabilities which have plagued adiabatic Chaplygin-gas models. The probability distribution for the deceleration parameter has a maximum at $q_{0} \approx -0.53$ which partially removes the degeneracy of previous studies which, taken at face value, were incompatible with an accelerated expansion and thus in obvious tension with results for the background.
With the matter fraction as a free parameter, our analysis also revealed that the matter fraction probability is highest for values smaller than roughly 8\%. This is a result in favor of the unified viscous model.
Consequently, as far as the matter power spectrum is concerned, the viscous model remains an option for a unified description of the dark sector. On the other hand, the simple viscous model, based on Eckart's theory, has problems to account for a gravitational potential that reproduces the CMB power spectrum. Possibly, this problem may be solved with the help of a causal transport theory for the bulk viscous pressure.

\section*{Acknowledgments}

Support by CNPq and FAPES is gratefully acknowledged.


\end{document}